\documentclass[showpacs,preprintnumbers,prd,showkeys,aps]{revtex4}

\usepackage{braket}
\usepackage{eurosym}
\usepackage{amssymb}
\usepackage{amsfonts}
\usepackage{amsmath}
\usepackage{graphicx}
\usepackage{calrsfs}
\usepackage{color}
\usepackage{subfig}
\usepackage[usenames,dvipsnames,svgnames,table]{xcolor}
\usepackage[colorlinks=true,
            linkcolor=red,
            urlcolor=red,
            citecolor=blue]{hyperref}

\begin{document}

\title{Quantum Corrected Schwarzschild Thin Shell Wormhole}
\author{Kimet Jusufi}
\email{kimet.jusufi@unite.edu.mk}
\affiliation{Physics Department, State University of Tetovo, Ilinden Street nn, 1200 Tetovo,
Macedonia}
\date{\today }

\begin{abstract}
Recently, Ali and Khalil \cite{Farag Ali}, based on the Bohmian quantum mechanics derived a quantum corrected version of the Schwarzschild metric. In this paper, we construct a quantum corrected Schwarzschild thin shell wormhole (QSTSW) and investigate the stability of this wormhole. First we compute the surface stress at the wormhole throat by applying the Darmois-Israel formalism to the modified Schwarzschild metric and show that exotic matter is required at the throat to keep the wormhole stable. We then study the stability analysis of the wormhole by considering phantom-energy for the exotic matter, generalized Chaplygin gas (GCG), and the linearized stability analysis. It is argued that, quantum corrections can affect the stability domain of the wormhole.
\end{abstract}
\pacs{04.20.Gz, 04.20.-q, 03.65.-w }
\keywords{Thin shell wormhole, Darmois-Israel formalism, Bohmian quantum mechanics, Stability }
\maketitle
\section{Introduction}

Morris and Thorne \cite{th1,th2}, showed that wormholes are solutions of Einstein field equations that connects two spacetime regions of the universe by a throat. Although their existence remains only of speculative nature, from a theoretical point of view, the models supporting their existence continues to be of great interest.  In particular, there are known two main difficulties that arise during their studies. Firstly, one needs to invoke the presence of the so-called exotic matter concentrated at the throat, which is shown to violate the weak energy condition (WEC), null energy condition (NEC), and Strong energy condition (SEC). Secondly, the problem of the stability of the wormhole, for such configurations to exists. On the other hand, Visser, introduced the concept of TSW \cite{visser1,visser2,visser3}, using Israel's junction conditions \cite{israel}. The basic idea behind Visser's method, is that by cutting and pasting two spacetime manifolds, one can construct a TSW which also minimizes the amount of the exotic matter at the wormhole throat. Furthermore, Poisson and Visser \cite{poisson}, investigated the stability of the STSW under a linearized radial perturbations around a static solution.  There are several models addressing the stability problem, such as: phantom-energy model, generalized Chaylang gas, linearized stability and many others. 

Different TSW solutions have been proposed, including, charged TSW \cite{ernesto,banerjee1}, TSW in heterotic string theory \cite{rahman}, TSW from noncommutative BTZ black hole \cite{banerjee2}, TSW in Einstein-Maxwell-Gauss-Bonnet gravity \cite{amirabi}, rotating TSW \cite{kashargin,mazh1}, TSW from scalar hair black hole \cite{ali}, TSW supported
by normal matter \cite{nm1,nm2}, cylindrical TSW \cite{sharifcy1,sharifcy2,zamorano,eiroacy,brannikovcy,peter2,bajarano}, and TSW with a cosmological constant \cite{lobo,kim},  wormholes in the framework of mimetic gravity \cite{myrzakulov} and references therein. Recently, Das \cite{das}, showed that by replacing the classical trajectories (geodesics) with the so-called quantal or Bohmian trajectories, gives rise to the quantum version of the Raychaudhuri equation (QRE). In particular, they also studied the QRE in the context of the Friedmann--Robertson--Walker (FRW) Universe, and argued that quantum corrections can prevents the formation of the big bang singularity \cite{das2}. The basic idea is to introduce a quantum velocity field $u_{\alpha}$, by writing
the wave function of a quantum fluid as \cite{Farag Ali,das,das2}
\begin{equation}
\psi(x^{\alpha})= \mathcal{R} e^{i\,S(x^{\alpha})},\label{1}
\end{equation}
where $\psi(x^{\alpha})$ is a normalizable wave function, $\mathcal{R}(x^{\alpha})$ and $S(x^{\alpha})$, are some real continuous function associated
with the four velocity field $u_{\alpha}=\frac{\hbar}{m} \, \partial_{\alpha}S$, in which $ (\alpha=0,1,2,3)$.  On the other hand, much less effort has been devoted to studies of the quantum effects in the context of the TSW. For example, in Ref. \cite{odintsov1,odintsov2} the possibility of primordial wormholes induced from GUTs  and in large N approximation at the early Universe was investigated. In Ref. \cite{sakalli}, the Hawking temperature from the traversable Lorentzian wormholes has been investigated. In Ref. \cite{lemos}, thermodynamics of rotating thin shells in the BTZ spacetime has been studied.  Motivated by what has been said, we aim to construct a QSTSW and investigate the effects of these quantum corrections on the stability of the wormhole. 

This paper is organized as follows. In Section 2, we start from the quantum corrected Schwarzschild metric and construct a QSTSW by computing the surface stress using the Darmois--Israel formalism. In Section 3, we study the effects of these corrections on the stability of the wormhole by considering first, the phantom-energy model for the exotic matter, then generalized Chaplygin gas (GCG), and finally the linearized stability around the static solution. In Section 4, we comment on our results.

\section{Quantum corrected Schwarzschild thin shell wormhole }

Recently, Ali and Khalil \cite{Farag Ali,das},  based on the Bohmian quantum mechanics found a quantum corrected Schwarzschild metric given by 
\begin{equation}
ds^{2}=-f(r)dt^{2}+f^{-1}(r)dr^{2}+r^{2}d\theta^{2}+ r^{2}\sin^{2}\theta d\phi^{2},\label{2}
\end{equation}
where 
\begin{equation}
f(r)=\left(1-\frac{2M}{r}+\frac{\hbar \eta}{r^{2}}\right).\label{3}
\end{equation}

Note that $\eta$ is a dimensionless constant, one can immediately see the analogy with the charged black hole metric simply by replacing $Q^{2}\to \hbar \eta$. It is interesting to note that, in contrast to the ordinary Schwarzschild solution, these corrections leads to a non-zero components of the stress-energy tensor $T_{\mu\nu}$. Morover, according to \cite{Farag Ali}, this may be a result of dark energy and dark matter. The outer horizon radius is calculated as
\begin{equation}
r_{h}=M\pm\sqrt{M^{2}-\eta \hbar}.\label{4}
\end{equation}

Furthermore, the corresponding Hawking temperature is shown to be
\begin{equation}
T_{H}=\frac{\sqrt{M^{2}-\hbar \eta}}{2\pi \left(M+\sqrt{M^{2}-\eta \hbar}\right)^{2}}.\label{5}
\end{equation}

Now we can use Cut and Paste technique to construct a QSTSW using the metric \eqref{2}. Therefore, let's consider now two copies of the above spacetime 
\begin{equation}
M^{(\pm)}=\left\lbrace r^{(\pm)}\geq a,\,\, a>r_{h} \right\rbrace,\label{6}
\end{equation}
and paste them at the boundary hypersurface $\Sigma^{(\pm)}= \left\lbrace r^{(\pm)}=a, a > r_{H}\right\rbrace$. This construction creates a geodesically complete manifold $M=M^{+}\bigcup M^{-}$. Next, by following the Darmois-Israel formalism, we write the original coordinates on $M$, as  $x^{\alpha} = (t,r, \theta, \phi)$, and the coordinates on the induced metric $\Sigma$, as $\xi^{i}=(\tau, \theta, \phi)$. Furthermore the parametric equation for $\Sigma$ is given by
\begin{equation}
\Sigma: F(r,\tau)=r-a(\tau)=0.\label{7}
\end{equation}

If we write the throat radius $a$, in terms of the the proper time $\tau$ on the
shell, $a=a(\tau)$, and then using the last equation it's not difficult to show that the induced metric on $\Sigma$ takes the form
\begin{equation}
ds^{2}_{\Sigma}=-d\tau^{2}+r^{2} \left(d\theta^{2}+\sin^{2}\theta \,d\phi^{2}\right).\label{8}
\end{equation}

On the other hand the Israel's junction conditions on $\Sigma$, reads
\begin{equation}
{S^{i}}_{j}=-\frac{1}{8 \pi}\left(\left[{K^{i}}_{j}\right]-{\delta^{i}}_{j}\,K\right).\label{9}
\end{equation}

Note that in the last equation, ${S^{i}}_{j}=diag(-\sigma, p_{\theta}, p_{\phi})$ is the energy momentum tensor on the thin-shell, $K$ and  $[K_{ij}]$, are defined as $K=trace\,[{K^{i}}_{i}]$ and  $[K_{ij}]={K_{ij}}^{+}-{K_{ij}}^{-}$, respectively. Morover the extrinsic curvature  ${K^{i}}_{j}$ is defined by
\begin{equation}
K_{ij}^{(\pm)}=-n_{\mu}^{(\pm)}\left(\frac{\partial^{2}x^{\mu}}{\partial \xi^{i}\partial \xi^{j}}+\Gamma^{\mu}_{\alpha \beta} \frac{\partial x^{\alpha}}{\partial \xi^{i}}\frac{\partial x^{\beta}}{\partial \xi^{j}}\right)_{\Sigma}.\label{10}
\end{equation}

In the last equation, $ {n_{\mu}}^{(\pm)}$, are the unit vectors normal to $M^{(\pm)}$ given as follows
\begin{equation}
n_{\mu}^{(\pm)}=\pm \left( \left\vert g^{\alpha \beta} \frac{\partial F}{\partial x^{\alpha}}\frac{\partial F}{\partial x^{\alpha}}\right\vert^{-1/2} \frac{\partial F}{\partial x^{\mu}}\right)_{\Sigma},     \label{11}
\end{equation}
such that $n_{\mu}n^{\mu}=1$. Using the above relations for the normal unit vectors to $M^{\pm}$ it's not difficult to show that
\begin{equation}
n_{\mu}^{(\pm)}=\left(\mp \dot{a}, \pm \frac{\sqrt{\dot{a}^{2}+f(a)}}{f(a)},0,0\right)_{\Sigma}.\label{12}
\end{equation} 

On the other hand, if we use Eq. \eqref{10} and Eq. \eqref{12}, for the extrinsic curvature components we find
\begin{eqnarray}
{K^{\phi}}^{(\pm)}_{\phi}&=&{K^{\theta}}^{(\pm)}_{\theta}=\pm \frac{1}{a}\sqrt{1-\frac{2M}{a}+\frac{\hbar \eta}{a^{2}}+\dot{a}^{2}},\label{13}\\
{K^{\tau}}^{(\pm)}_{\tau}&=&\pm \frac{\frac{M}{a^{2}}-\frac{\hbar \eta}{a^{3}}+\ddot{a}}{ \sqrt{1-\frac{2M}{a}+\frac{\hbar \eta}{a^{2}}+\dot{a}^{2}}}.\label{14}
\end{eqnarray}

Using the above results, one can easily check that the surface density and the surface pressure are given by the following relations
\begin{eqnarray}
\sigma &=&-\frac{1}{2 \pi a  }\sqrt{1-\frac{2M}{a}+\frac{\hbar \eta}{a^{2}}+\dot{a}^{2}},\label{15}\\
p&=&p_{\theta}=p_{\phi}=\frac{1}{4 \pi a}\frac{1-\frac{M}{a}+\dot{a}^{2}+a \dot{a}}{\sqrt{1-\frac{2M}{a}+\frac{\hbar \eta}{a^{2}}+\dot{a}^{2}}}.\label{16}
\end{eqnarray}

From the last two equations we can now write the static configuration of radius $a$, by setting $\dot{a}=0 $, and $ \ddot{a}=0$, we get
\begin{eqnarray}
\sigma_{0} &=&-\frac{1}{2 \pi a_{0}  }\sqrt{1-\frac{2M}{a_{0}}+\frac{\hbar \eta}{a^{2}_{0}}},\label{17}\\
p_{0}&=&\frac{1}{4 \pi a_{0}}\frac{1-\frac{M}{a_{0}}}{\sqrt{1-\frac{2M}{a_{0}}+\frac{\hbar \eta}{a^{2}_{0}}}}.\label{18}
\end{eqnarray}

From Eq. \eqref{17} one can see that the surface density is always negative, i.e. $\sigma_{0} <0$, as a consequence of this, the (WEC) is violeted. Furthermore, one can easly check that also the (NEC) is violated, i.e. $\sigma_{0}+p_{0}<0$.  Now, one can calculate the amount of exotic matter needed to construct the wormhole by solving the following integral
\begin{equation}
\Omega_{\sigma}=\int \sqrt{-g}\,\left(\rho+p_{r}\right)\,d^{3}x.\label{19}
\end{equation}

For a thin-shell wormhole $p_{r}=0$ and $\rho=\sigma \delta(r-a)$, where $\delta(r-a)$ is the Dirac delta function. Solving the above integral and inserting the value of surface energy density, in static configuration we find
\begin{equation}
\Omega_{\sigma}=\int_{0}^{2\pi}\int_{0}^{\pi}\int_{-\infty}^{\infty}\sigma \sqrt{-g}\,\delta(r-a)dr\,d\theta\,d\phi.\label{20}
\end{equation}

Since the shell does not exert radial pressure and the energy density is
located on a thin shell surface, it follows that
\begin{equation}
\Omega_{\sigma}=-2a_{0}\sqrt{1-\frac{2M}{a_{0}}+\frac{\hbar \eta}{a_{0}^{2}}}.\label{21}
\end{equation}

Let us now analyze the attractive and repulsive
nature of the wormhole. Starting from the observer’s four-acceleration given as,
$a^{\mu}=u^{\nu}\nabla_{\nu}u^{\mu}$, where the four velocity reads $u^{\mu}=(1/\sqrt{f(r)},0,0,0)$. It's not difficult to show that we are left only with the contribution of the radial component 
\begin{equation}
a^{r}=\Gamma^{r}_{tt}\left(\frac{dt}{d\tau}\right)^{2}=\frac{M}{r^{2}}-\frac{\hbar \eta}{r^{3}}.\label{22}
\end{equation}

As a consequence of the last equation, a test particle obeys the equation of motion
\begin{equation}
\frac{d^{2}r}{d\tau^{2}}=-\Gamma^{r}_{tt}\left(\frac{dt}{d\tau}\right)^{2}=-a^{r}.\label{23}
\end{equation}

Note that this equation gives the geodesic equation if $a^{r}=0$.  Also, we
observe that the wormhole is attractive when $a^{r}>0$ and
repulsive when a
$a^{r}<0$. For more useful information on the effects of quantum corrections on the acceleration of the wormhole we show graphically in Figure 1.

\begin{figure}
\centering
\begin{minipage}{.45\textwidth}
\centering
\includegraphics[width=.95\linewidth]{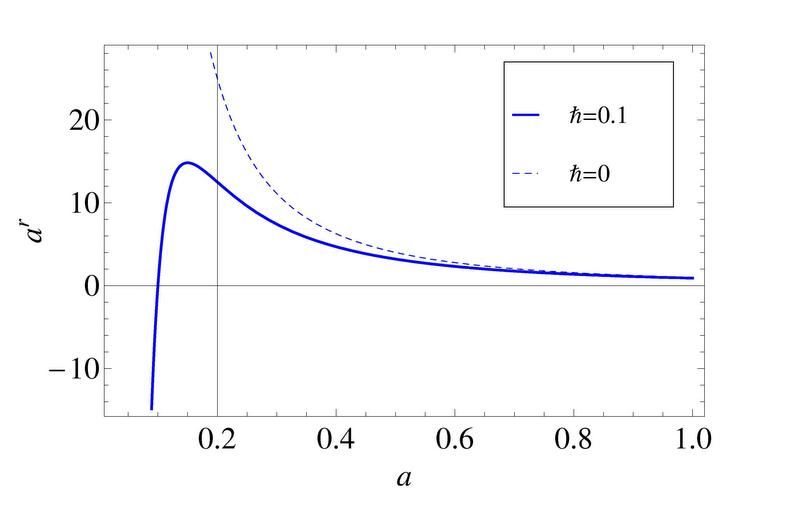}
\caption{\small \textit{ Plots for the acceleration of the QSTSW/STSW corresponding to $\eta=m=1$ and $\hbar=0.1$, $\hbar=0$, respectively.}}
\end{minipage}
\begin{minipage}{.45\textwidth}
\centering
\includegraphics[width=.9\linewidth]{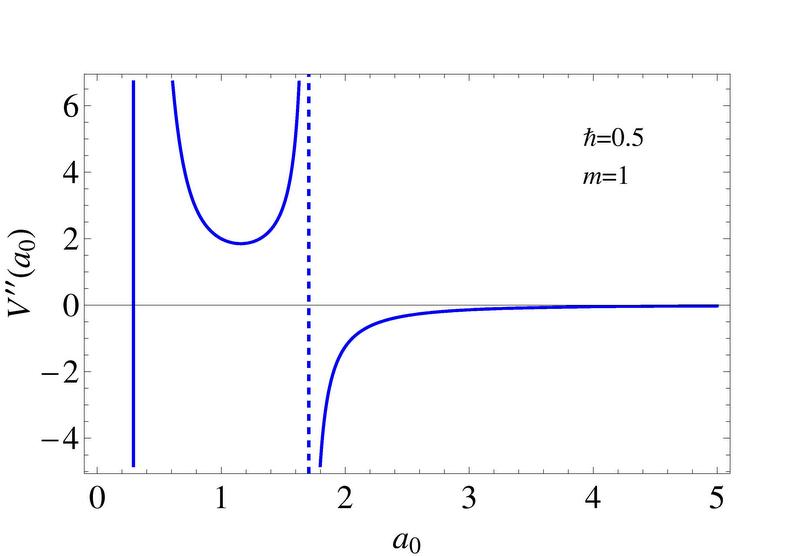}
\caption{ \small \textit{Stability regions of QSTSW corresponding to $\eta=m=1$ and $\hbar=0.5$.  The dashed vertical line corresponds to $r_{h}$, the regions at the left of the dashed vertical line have no physical meaning $(a_{0}\leq r_{h})$.}}
\end{minipage}
\end{figure}

\section{Stability Analysis}
\subsection{Phantom-like equation of state}

We will now analyze the stability of the shell by considering the phantom-like equation of state given as  $p=\omega \sigma$, where $\omega<0$ \cite{peter1}. From the conservation equation, one can check that the surface pressure and energy density obey the following relation

\begin{equation}
\nabla_{j}S^{ij}=\frac{d\sigma}{d\tau}+\frac{2 \dot{a}}{a}\left(\sigma+p\right)=0,\label{24}
\end{equation}
now we make use of equation of state $p=\omega \sigma$, yielding
\begin{equation}
\frac{d\sigma}{da}+\frac{2 \sigma}{a}\left(1+\omega\right)=0.\label{25}
\end{equation}

Solving the last equation for $\sigma$, we find
\begin{equation}
\sigma(a)=\sigma(a_{0})\left(\frac{a_{0}}{a}\right)^{2(1+\omega)},\label{26}
\end{equation}
in which we have used  $\sigma_{0}= \sigma(a_{0})$. Equating this result with the Eq. \eqref{15}, leds to the equation of motion
\begin{equation}
\dot{a}^{2}+V(a)=0\label{27}
\end{equation}
where the potential reads
\begin{equation}
V(a)=f(a)-4 \pi^{2}a^{2}\sigma^{2}.\label{28}
\end{equation}

On the other hand if we substitute Eq. \eqref{26} into Eq. \eqref{28} we get
\begin{equation}
V(a)=1-\frac{2M}{a}+\frac{\hbar \eta}{a^{2}}-4\pi^{2}\sigma_{0}^{2}a^{2}\left(\frac{a_{0}}{a}\right)^{4(1+\omega)}.\label{39}
\end{equation}

Since we aim to perturb the throat of the thin-shell wormhole radially around the equilibrium radius $a_{0}$, we can use Taylor's series by expanding the potential around $a=a_{0}$, and get
\begin{equation}
V(a)=V(a_{0})+V^{\prime}(a_{0})(a-a_{0})+\frac{1}{2}V^{\prime\prime}(a_{0})(a-a_{0})^{2}+\mathcal{O}(a-a_{0})^{3}\label{30}.
\end{equation}

One can now esaly check that $V(a_{0})=0$. Furthermore, we also require that $V^{\prime}(a_{0})=0$, therfore solving for $\omega$ we find
\begin{equation}
\omega=-\frac{a_{0}\left(a_{0}-M\right)}{2\left(a_{0}^{2}-2a_{0}M+\eta \hbar \right)}.\label{31}
\end{equation}

Now we can go back and substitute the result for $\omega$ into the Eq. \eqref{30}, which for static configuration  gives 
\begin{equation}
V^{\prime\prime}(a_{0})=-\frac{2}{a_{0}^{3}}\frac{\left(a_{0}^{2}M-2a_{0}\eta \hbar+M\eta \hbar \right)}{\left(a_{0}^{2}-2a_{0}M+\eta \hbar \right)}.\label{32}
\end{equation}

The QSTSW solution is stabile if and only if $V^{\prime\prime}(a_{0})>0$,  therefore, one can show that the following inequality holds
\begin{equation}
\frac{\sqrt{\hbar \eta}}{M}>\frac{a_{0}/M}{\sqrt{2 a_{0}/M-1}}.\label{33}
\end{equation}

To meet this condition, $\hbar$, needs to exceed $M$, we can therefore conclude that, the wormhole is unstable. Note that we found analogues results to the case of charged thin shell wormhole by replacing $Q^{2}\to \hbar \eta$ \cite{peter1}. This result can be seen more clearly in Figure 2.

\subsection{Stability by Generalized Chaplygin gas}

Another interesting model dealing with the stability of the TSW is the generalised Chaplygin gas. According to this model, the exotic matter at the throat can be modeld by the following equation of state \cite{amirabi,lobo2006}
\begin{equation}
p=\left(\frac{\sigma_{0}}{\sigma}\right)^{\gamma}\sigma,\label{34}
\end{equation}
where $0<\gamma \leq1$, $\sigma$ is surface energy density and $p$ is surface pressure. If we substitute the last equation into the Eq. \eqref{24}, and solve for $\sigma$, we find the following result
\begin{equation}
\sigma(a)=\sigma_{0}\left[\left(\frac{a_{0}}{a}\right)^{2(1+\gamma)}+\frac{p_{0}}{\sigma_{0}} \left(\left(\frac{a_{0}}{a}\right)^{2(1+\gamma)}-1\right)\right]^{\frac{1}{1+\gamma}}.\label{35}
\end{equation}

Equating this result with Eq. \eqref{15}, it follows the equation of motion
\begin{equation}
\dot{a}^{2}+V(a)=0,\label{36}
\end{equation}
in which
\begin{equation}
V(a)=f(a)-4\pi^{2}a^{2}\sigma_{0}^{2}\left[\left(\frac{a_{0}}{a}\right)^{2(1+\gamma)}+\frac{p_{0}}{\sigma_{0}} \left(\left(\frac{a_{0}}{a}\right)^{2(1+\gamma)}-1\right)\right]^{\frac{2}{1+\gamma}}.\label{37}
\end{equation}

At the static configuration, i.e. $a=a_{0}$, from the last equation, $V(a_{0})$ and $V^{\prime}(a_{0})$, are shown to vanish, i.e. $V(a_{0})=V^{\prime}(a_{0})=0$. We can now use the result of Taylor's expansion of $V(a)$ about
$a = a_{0}$,  to find 
\begin{equation}
V^{\prime\prime}(a_{0})=-\frac{2\left[a_{0}(1+\gamma)(3M^{2}+a_{0}^{2})+2a_{0}\gamma \eta \hbar-M\left(a_{0}^{2}(3+4\gamma)+(1+2\gamma)\eta \hbar \right)\right]}{a_{0}^{3}\left(a_{0}^{2}-2Ma_{0}+\eta \hbar \right)}.\label{38}
\end{equation}

We know that the static configuration of the TSW is in stable equilibrium if $ V^{\prime\prime}(a_{0})>0$. Solving the last equation for $\gamma$ we find 
\begin{equation}
\gamma<\frac{3Ma_{0}^{2}-3M^{2}a_{0}-a_{0}^{3}+M\eta \hbar}{\left(M-a_{0}\right)\left(3Ma_{0}-a_{0}^{2}-2\eta \hbar \right)}.\label{39}
\end{equation}

From the horizon radius $r_{h}$ it follows that quantum effects $\eta \hbar$, cannot exceed the black hole mass $M$, i.e. $M^{2}\geq \eta \hbar$. Therefore, we will only consider the stability domain of the wormhole in the following interval $0<\sqrt{\hbar\, \eta}/M \leq 1$. From Eq. \eqref{39} if follows that, there exists some part of the parameter space where the throat location is stable. However, one can convince himself that, if we increase the quantum effects by using different values of $\hbar$, the stability domain also increases. For more useful information we show the stability region graphically in Figure 3.

\begin{figure}[h!]
\includegraphics[width=0.40\textwidth]{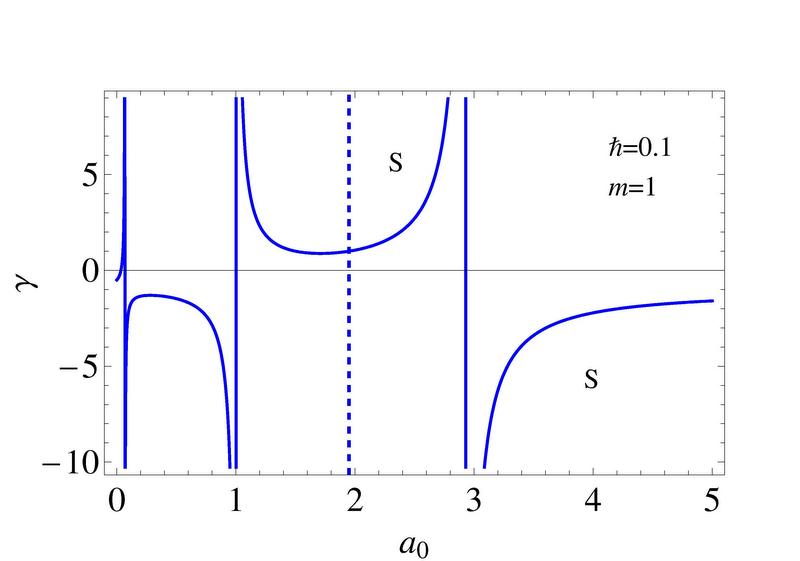} %
\includegraphics[width=0.40\textwidth]{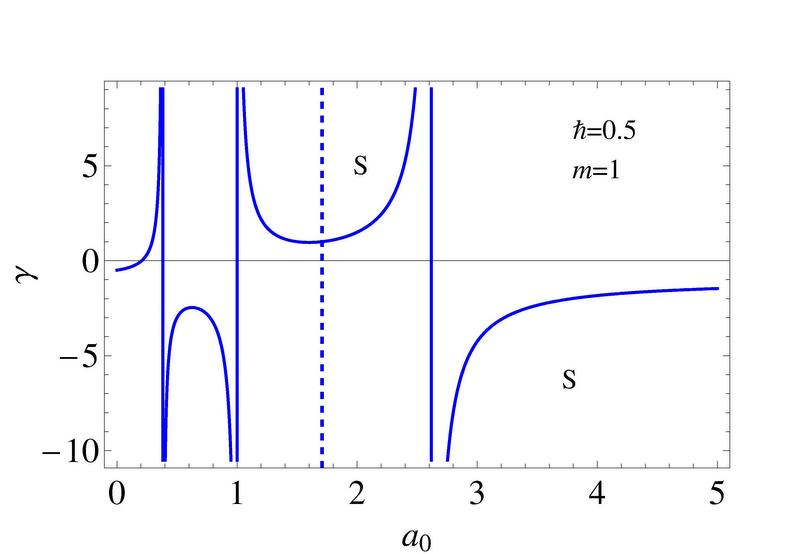} %
\includegraphics[width=0.40\textwidth]{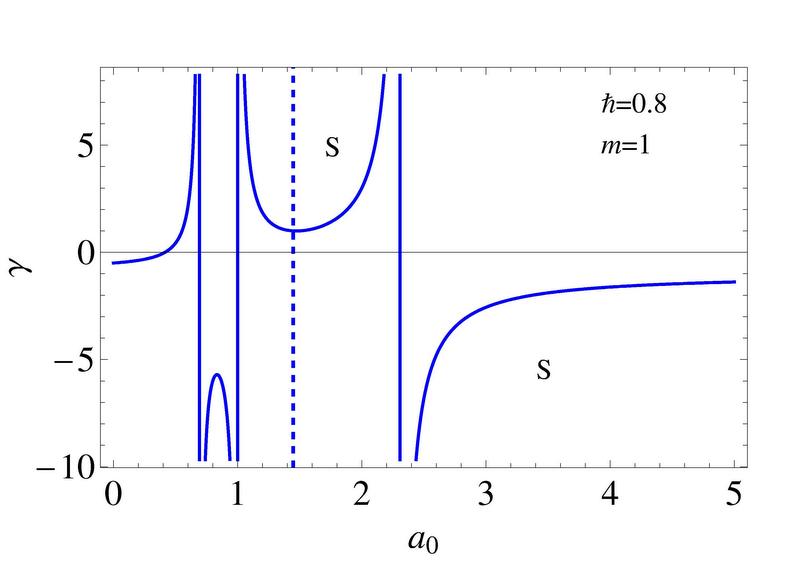} %
\includegraphics[width=0.40\textwidth]{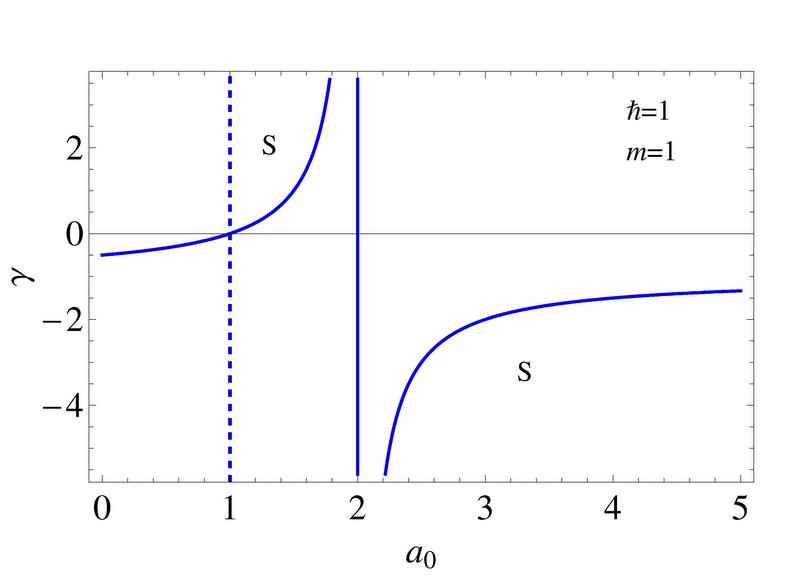}
\caption{\small \textit{Stability regions of QSTSW  in terms of $\gamma$ and radius of the throat $a_{0}$ for $\eta=m=1$  and different values of $\hbar$. The stable
regions are denoted by S and are situated at the right of the dashed vertical lines which corresponds to $r_{h}$.} }
\end{figure}

\subsection{Linearized stability analysis}

Finally, here we will try to analyze the stability of quantum corrected TSW by considering the linearized stability analysis. As we have seen, performing linear perturbations around $a=a_{0}$, gives
\begin{equation}
V(a)=V(a_{0})+V^{\prime}(a_{0})(a-a_{0})+\frac{1}{2}V^{\prime\prime}(a_{0})(a-a_{0})^{2}+\mathcal{O}(a-a_{0})^{3}.\label{40}
\end{equation} 

Simply by rearranging \eqref{17}, once again we get the equation of motion 
\begin{equation}
\dot{a}^{2}+V(a)=0,\label{41}
\end{equation}
with the potential 
\begin{equation}
V(a)=f(a)-4 \pi^{2}a^{2} \sigma^{2}.\label{42}
\end{equation}

On the other hand we can rewrite Eq. \eqref{17} as follows
\begin{equation}
\sigma{\prime}=-\frac{2}{a}\left(\sigma+p\right)=0,\label{43}
\end{equation}
where the prime means $d/d a$. Let us now use this result together with Eq. \eqref{42} and find that the first derivative of the potential is given by
\begin{equation}
V^{	\prime}(a)=\frac{2M}{a^{2}}-\frac{2\hbar \eta}{a^{3}}+8 \pi^{2} a \,\sigma \left(\sigma+2p\right).\label{44}
\end{equation}

We can follow the work in \cite{poisson}, in which the equation of state of the matter
which supports the TSW was chosen to be $p=p(\sigma)$. Therefore, we can define a parameter $\beta$ by the relation
\begin{equation}
\beta^{2}(\sigma)=\frac{dp}{d\sigma},\label{45}
\end{equation}
which for ordinary matter is interpreted as the velocity of sound. Now if we set $a=a_{0}$, and due to the linearization we have $V(a_{0})=V^{\prime}(a_{0})=0$, therefore, we are left with the following result
\begin{equation}
V^{\prime\prime}(a_{0})=-\frac{2 \left[a_{0}^{4}(1+2\beta^{2}_{0})-a_{0}^{3}M(3+10 \beta^{2}_{0})-a_{0}M (1+14\beta^{2}_{0})\eta \hbar +4 \beta^{2}_{0}\eta^{2}\hbar^{2}+3a_{0}^{2}(M^{2}(1+4\beta^{2}_{0})+2\beta^{2}_{0}\eta \hbar)\right]}{a_{0}^{4}\left(a_{0}^{2}-2a_{0}M+\eta \hbar\right)}.\label{46}
\end{equation}

Note that we have used $\beta_{0}^{2}=\beta^{2}(\sigma_{0})$ and should satisfy $0<\beta^{2}_{0}\leq 1$. The stability requirement for the wormhole solution is written as $V^{\prime\prime}(a_{0})>0$. Solving for $\beta_{0}^{2}$ we find
\begin{equation}
\beta_{0}^{2}<\frac{3a_{0}^{3}M-3a_{0}^{2}M^{2}+a_{0}M\eta \hbar-a_{0}^{4}}{2a_{0}^{4}-10a_{0}^{3}M+12a_{0}^{2}M^{2}+6a_{0}^{2}\eta \hbar-14 a_{0}M\eta  \hbar+4\eta^{2}\hbar^{2} \hbar}.\label{47}
\end{equation}

This means there exists some part of the parameter space where the throat location is stable. For more useful information we show the stability region graphically in Figure 4. 

As we can see, quantum corrections increases the possibility of obtaining stable thin-shell wormhole in the same way as charge. As we know in the case of a charged black hole the new energy source corresponds to the contribution of the electromagnetic field encoded in the stress--energy tensor. In a similar way here, it may be possible that, these quantum effects to be related with the presence of a new form of energy encoded in the stress--energy tensor. More specifically, by solving the Einstein field equations $G_{\mu\nu}=8\pi T_{\mu\nu}$ we can derive the non--zero values of the stress--energy tensor $T_{\mu\nu}$ (see, e.g., Eq. (26) in Ref. \cite{Farag Ali}). The origin of this energy is unclear, however, according to \cite{Farag Ali} this energy may be related to dark matter and dark energy. If this is the case, then our results suggest that wormholes are supported by the presence of dark matter and dark energy. Finally, the results found in this paper are consistent with the results found in Ref. \cite{odintsov1,odintsov2} where it was argued that quantum effects increases the possibility of obtaining stable primordial wormholes at the early Universe.

\begin{figure}[h!]
\includegraphics[width=0.40\textwidth]{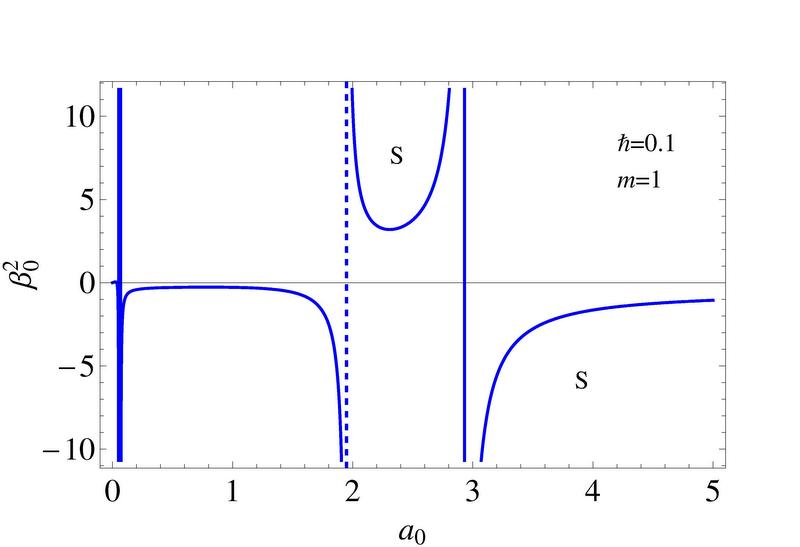} %
\includegraphics[width=0.40\textwidth]{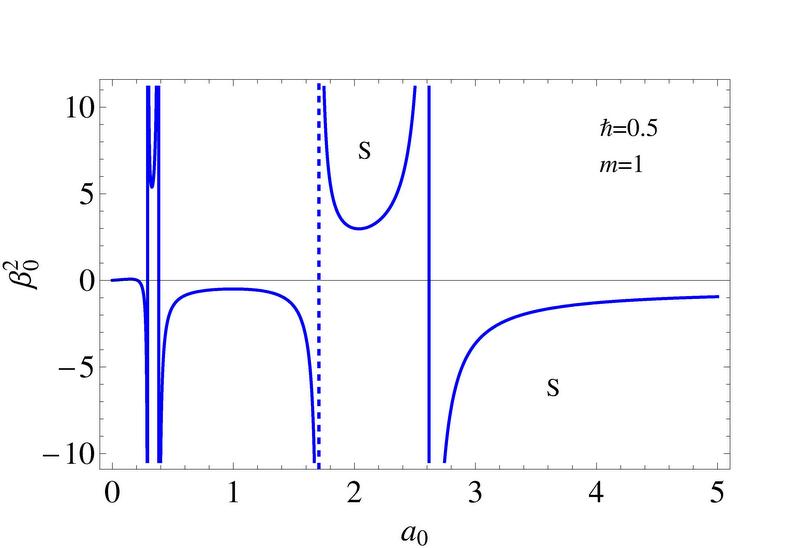} %
\includegraphics[width=0.40\textwidth]{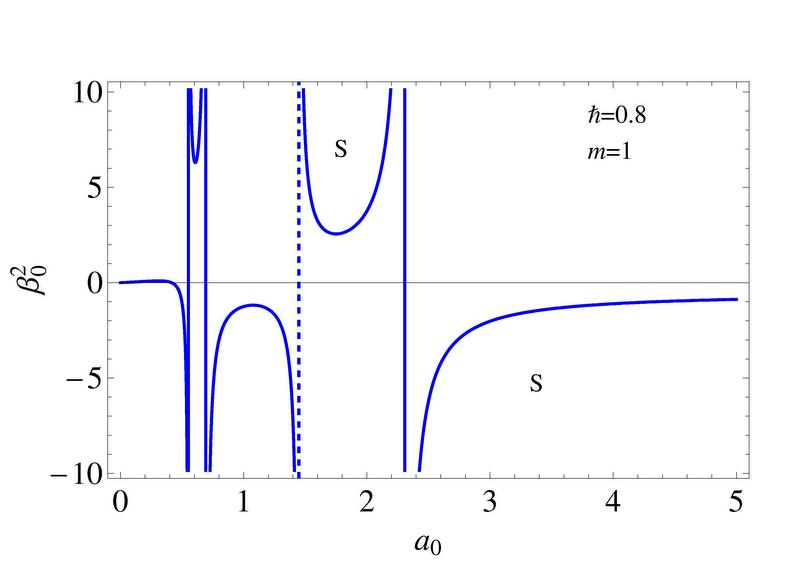} %
\includegraphics[width=0.40\textwidth]{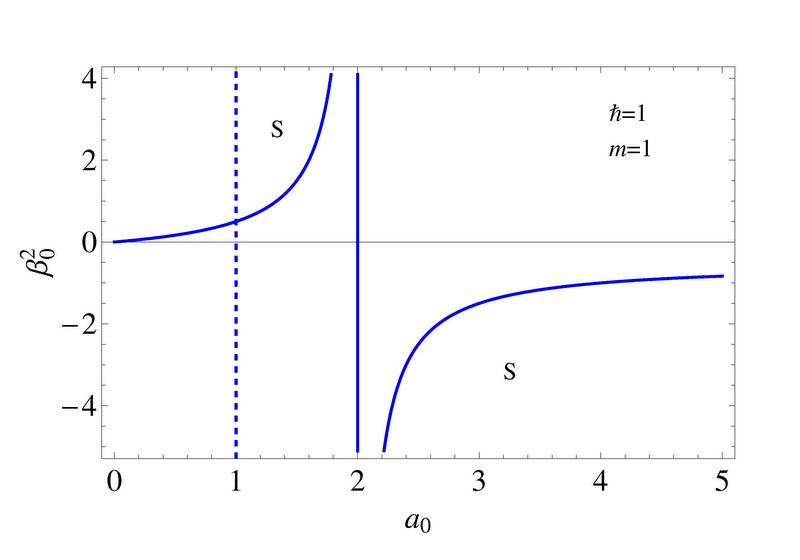}
\caption{ \small \textit{
Stability regions of QSTSW in terms of $\beta^{2}_{0}$ and radius of the throat $a_{0}$, for $\eta=m=1$, and different values of $\hbar$. The stable
regions are denoted by S and are situated at the right of the dashed vertical lines which corresponds to $r_{h}$.}}
\end{figure}

\vspace{1cm}
\section{Conclusion}
In this work, we have constructed a QCTSW by using Visser's method and shown that exotic matter is required at the throat to keep the wormhole stable.  The stability analyses first is carried out by using the Phantom-like equation of state $p=\omega \sigma$, and shown the TSW to be unstable since it required $\hbar$ to exceed $M$.  However, using the methods of generalized Chaplygin gas and linearized stability analysis, we show that the wormhole can be stable by choosing suitable values of parameters. More specifically, by choosing different values of $\hbar$ in the  following interval $0<\sqrt{\hbar\, \eta}/M \leq 1$ it is shown that, quantum corrections increases the stability domain of obtaining stable wormhole solutions. Note that, the role of quantum corrections on the stability of the wormhole is similar to the role played by charge on the wormholes stability \cite{ernesto}. In the end we speculate that, if the origin of these quantum effects is related to the presence of dark matter and dark energy as suggested by \cite{Farag Ali}, the bottom line of this reasoning is that dark matter and dark energy may support the stability of the wormhole.

\end{document}